\documentclass[12pt]{iopart}
\usepackage{graphicx}

\begin{document}

\title{Cold atoms probe the magnetic field near a wire}

\author{M P A Jones, C J Vale\footnote{Current address: Dept. of Physics,
University of Queensland, Brisbane 4072, Australia.}, D Sahagun, B
V Hall, C C Eberlein\dag, B E Sauer, K Furusawa$^\ast$, D
Richardson$^\ast$ and E A Hinds}
\address{Blackett Laboratory, Imperial College, London SW7 2BW, UK}
\address{\dag Department of Physics and Astronomy, Sussex University, Falmer, BN1 9QH,UK}
\address{$^\ast$ Optoelectronics Research Centre, Southampton University, Southampton, SO17 1BJ, UK}
\ead{ed.hinds@ic.ac.uk}

\begin{abstract}
A microscopic Ioffe-Pritchard trap is formed using a straight,
current-carrying wire, together with suitable auxiliary magnetic
fields.  By measuring the distribution of cold rubidium atoms held
in this trap, we detect a weak magnetic field component $\Delta
B_z$ parallel to the wire. This field is proportional to the
current in the wire and is approximately periodic along the wire
with period $\lambda=230\,\mu$m. We find that the decrease of this
field with distance from the centre of the wire is well described
by the Bessel function $K_1(2\pi y/\lambda)$, as one would expect
for the far field of a transversely oscillating current within the
wire.
\end{abstract}


 \maketitle

\section{Introduction}

The ability to control cold atom clouds in microscopic magnetic
traps \cite{weinstein95,vuletic98,fortagh98} and waveguides
\cite{mueller99,dekker00,key00} has created the  new field of
miniaturized atom optics \cite{hindsreview99,folmanreview02}. With
the use of microstructured surfaces (atom chips) it becomes
possible to control cold atoms on the $\mu$m length scale and to
anticipate the construction of integrated atom interferometers
\cite{hinds01,haensel01,andersson02}. Ultimately there is the
possibility of controlling the quantum coherences within arrays of
individual atoms for use in quantum information processing
\cite{calarco00,horak02}. For these kinds of applications it is
important to avoid fluctuating or inhomogeneous perturbations,
which tend to destroy the quantum coherences.

Cold atom clouds cooled below a few $\mu$K have recently been used
to probe the magnetic field fluctuations within 100\,$\mu$m of a
current-carrying wire.  Two frequency domains have been studied.
Audiofrequency fluctuations of the currents that form the
microtraps can excite centre-of-mass vibrations of the atoms
\cite{haensel01,fortagh02}, causing the cloud to heat up.
Radiofrequency noise in the wire currents can drive spin flips,
which cause atoms to be ejected from the trap
\cite{haensel01,fortagh02}. Usually this noise is due to technical
imperfections of the apparatus, as elucidated by Leanhardt {\it et
al.} through a comparison of magnetic and optical traps near a
surface \cite{leanhardt02a}. However, there is also a fundamental
component of the magnetic field noise in the near-field of the
wire, which is due to thermal fluctuations of the current and has
recently been measured in our laboratory \cite{jones03}.

The dc behaviour of the field is less well understood.  As the
atom cloud is brought close to the main current-carrying wire of
the trap it breaks into fragments along the length of the wire
 \cite{fortagh02,leanhardt02,jones03}, indicating
an unexpected variation in the trapping potential. Kraft {\it et
al.}\cite{kraft02} have recently shown that this is due to the
presence of a magnetic field component $\Delta B_z$ {\it parallel}
to the wire that varies along the length of the wire. In this
paper we use cold atoms held in a microtrap close to a wire to
measure $\Delta B_z$ and to explore how it varies in the vicinity
of the wire.

\section{Preparation of cold atoms}

\begin{figure}
\centering
\includegraphics[width=3.3in]{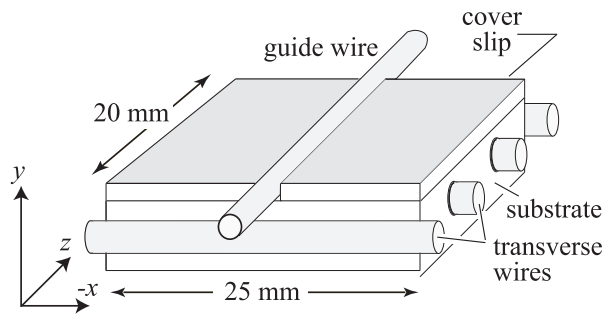}
\caption{Construction of the magnetic microtrap (not to scale). A
current along $\it{z}$ and a bias field along $\it{x}$ together
form a long thin trap parallel to the guide wire.  The ends of the
trap are closed by currents in the transverse wires.  Atoms are
trapped near the surface of the guide wire.} \label{fig:chip}
\end{figure}

The arrangement of wires used to form our microtrap is shown in
figure \ref{fig:chip}. The main wire is a 500\,$\mu$m diameter
guide wire along the {\it z}-direction. In cross section it has a
$370\,\mu$m diameter copper core, surrounded by an aluminium layer
$55\,\mu$m thick with a $10\,\mu$m thick ceramic outer coating.
This wire is glued by high-vacuum epoxy (Bylapox 7285) into the
$200\,\mu$m-deep channel formed by a glass substrate and two glass
cover slips. Below the guide wire there are four transverse wires,
each $800\,\mu$m in diameter. The cover slips are coated with
60\,nm of gold so that they reflect 780\,nm light. In order to
load cold atoms into the microtrap we first collect $^{87}$Rb
atoms using a magneto-optical trap whose beams are reflected from
the gold surface. This MOT collects $1\,\times\,10^8$ atoms at a
height of 4\,mm above the surface and cools them to $50\,\mu$K.
The MOT is pulled down to a height of 1.3\,mm by passing a current
of 3.2\,A through the guide wire and adding a uniform magnetic
field $B^{bias}_x$ of 6\,G along the {\it x}-direction.  This
compresses the cloud into a cylindrical shape and increases the
phase space density of the atoms to $2 \times 10^{-6}$
\cite{jones02}.

The light and the anti-Helmholtz coils of the MOT are then
switched off and the atoms are optically pumped into the $|F,m
\rangle = |2,2 \rangle$ state. We collect $2\,\times\,10^7$ of
these atoms in the magnetic guide formed by the guide wire (8\,A
along {\it z}) and the transverse bias field $B^{bias}_x$ (10\,G
along {\it x}). Axial confinement is provided by the inner
transverse wires (15\,A each along {\it -x} ) and the outer
transverse wires (15\,A each along {\it x} ). The field at the
centre of this trap is partly cancelled by an axial bias field
$B^{bias}_z$ ($6\,$G along {\it z}). Next, the trap is
adiabatically compressed over 0.5\,s by increasing $B^{bias}_x$
and $B^{bias}_z$ to $29\,$G and $11\,$G respectively, and reducing
the guide current to $6.9\,$A. This brings the trap to a height of
225\,$\mu$m above the top of wire and reduces the net field along
$z$ at the centre of the trap to $\sim 1\,$G. In this way, the
radial and axial trap frequencies are raised to 840\,Hz and
26\,Hz. The elastic collision rate is now $\sim 54$\,s$^{-1}$,
which is high enough for forced rf evaporative cooling to be
efficient. We sweep the rf frequency logarithmically over 6\,s
from 13\,MHz to a final frequency near 2.8\,MHz, cooling the cloud
down to approximately 6\,$\mu$K. This temperature is a good choice
because it gives the atoms a thermal kinetic energy comparable
with the potential created by $\Delta B_z$ that we wish to probe.
The atoms are brought closer to the surface by a smooth reduction
of the current flowing in the guide wire during the last $1\,$s of
evaporation. The arrival of the atoms at the desired height
coincides with the end of the evaporation ramp.

\begin{figure}
\centering
\includegraphics[width=3.4in]{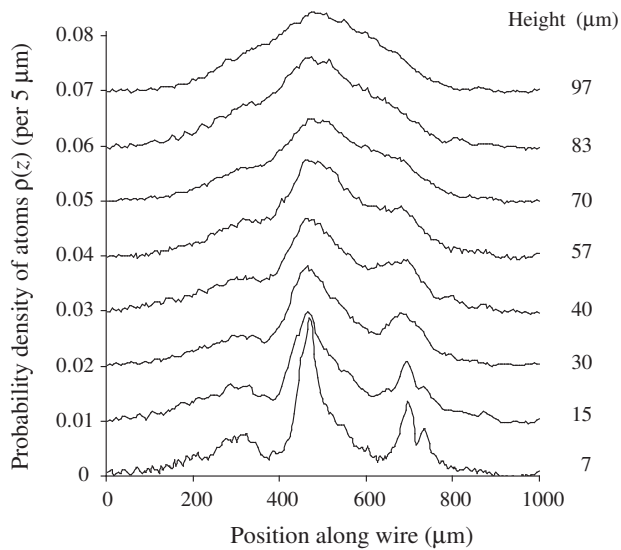}
\caption{Distribution of atoms $\rho(z)$ along the length of the
Ioffe-Pritchard trap for a variety of distances between the atoms
and the guide wire. The curves show the probability of finding
atoms within a range $\delta z=5\mu$m and are all normalized to
unit total probability. They are successively offset by 0.01 for
the sake of clarity. Far from the wire, the distribution is
Gaussian, but at closer distances the cloud develops additional
structure and can break into lumps} \label{fig:curves}
\end{figure}

The cloud is viewed using a ccd camera to record the absorption of
a resonant probe beam propagating along the {\it x}-direction. Two
images are recorded, one with the atoms present and one without,
and we compute the logarithm of their ratio. Since the cloud is
optically thin, this procedure yields an image of the column
density of the atoms, viewed in the $y-z$ plane. Integrating this
over $y$, gives the probability distribution of atoms along the
{\it z}-direction. Let us call it $\rho(z)$. Assuming that the gas
is in thermal equilibrium at temperature $T$, $\rho(z)$ is related
to the confining potential $U(z)$ by the Boltzmann factor
$\exp(-U(z)/kT)$.  In the absence of $\Delta B_z$, the potential
$U(z)$ is harmonic with an axial frequency of 26-27\,Hz over the
range of heights studied. Fitting a parabola to $-\ln\rho(z)$ then
determines the temperature $T$. For example, the top curve in
figure \ref{fig:curves} taken 97\,$\mu$m above the surface of the
wire gives $T=5.8\,\mu$K.

\section{Fragmentation}

When the cloud is brought closer to the surface, the distribution
$\rho(z)$ develops additional structure, as can be seen in figure
\ref{fig:curves}. Similar behaviour has been reported by other
groups studying atoms near copper wires
\cite{fortagh02,leanhardt02}. This indicates that the atoms
experience some other interaction near the surface in addition to
the expected simple harmonic potential of the trap. Kraft {\it et
al.} have recently shown that this is due to a magnetic field
component $\Delta B_z$ {\it parallel} to the wire \cite{kraft02}.
Where this field adds to (reduces) the axial bias field of the
magnetic trap there is an increase (decrease) of $\mu_B \Delta
B_z$ in the potential energy of the trapped atoms. To obtain
$\Delta B_z$ for each curve in figure \ref{fig:curves}, we first
derive the temperature $T$ from the parabola fitted to
-ln$\rho(z)$. This parabola is chosen to pass through the centre
of the $\Delta B_z$ oscillations.  For the 8 curves shown the
temperature is nearly constant, being (5.8, 6.1, 6.4, 7.0, 7.8,
7.8, 7.8, 7.8)\,$\mu$K, with the variations arising from slight
differences in the final stage of the evaporation trajectories.
The difference between $-\ln\rho(z)$ and the fitted harmonic
potential is the potential of interest $\Delta B_z/kT$, and since
we know $T$ from the harmonic fit, we obtain $\Delta B_z$. In this
way we have measured how $\Delta B_z$ varies along $z$ at the 8
different heights $h$ of the cloud above the wire.

\begin{figure}
\centering
\includegraphics[width=4in]{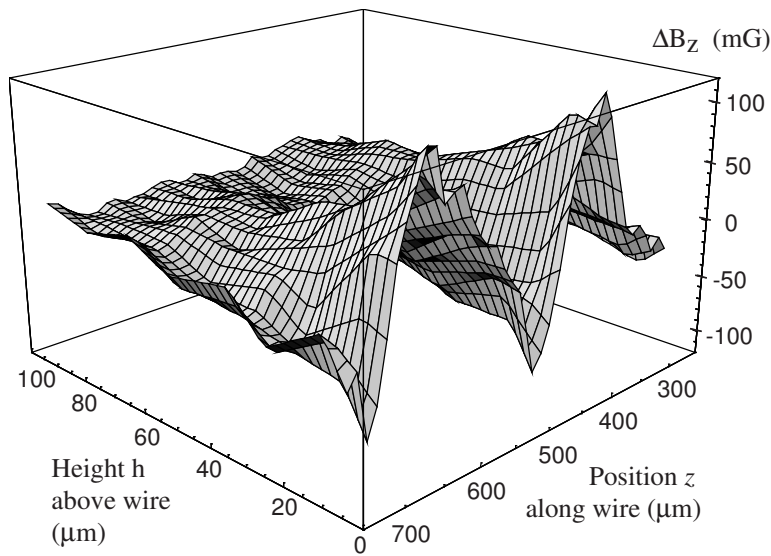}
\caption{Anomalous magnetic field $\Delta B_z$ as a function of
position for a current of 3.7\,A in the guide wire.}
 \label{fig:potential}
\end{figure}

The height of the cloud above the surface is set by adjusting the
current in the guide wire, keeping the bias field $B^{bias}_x$
fixed at 29\,G. For the largest and smallest heights presented in
figure \ref{fig:curves} (97\,$\mu$m and 7.2\,$\mu$m) the guide
wire currents are 5\,A and 3.7\,A respectively. $\Delta B_z$ is
proportional to the current in the guide wire \cite{kraft02}, so
to compare $\Delta B_z$ at different heights we scale the result
at each height to the current used. Figure \ref{fig:potential} is
a map of $\Delta B_z(y,z)$ scaled to a fixed 3.7\,A in the guide
wire. The surface shown interpolates between the 8 heights where
measurements were made.

\begin{figure}
\centering
\includegraphics[width=4in]{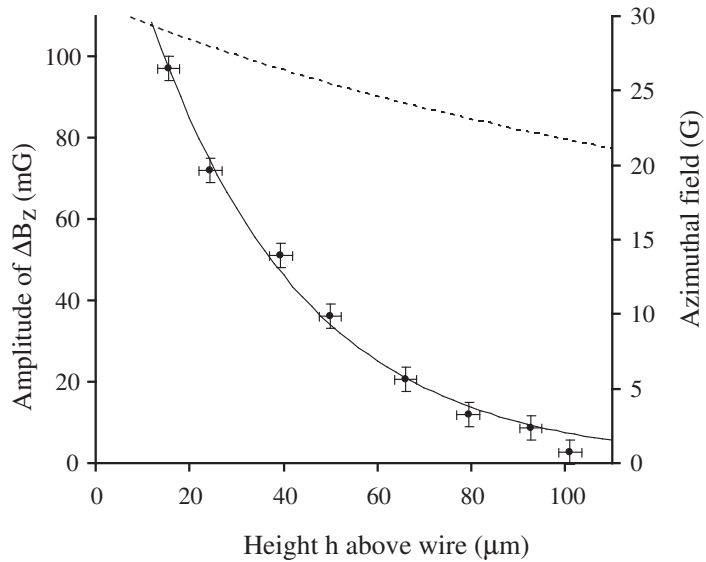}
\caption{Magnetic fields versus height $h$ above a wire carrying
3.7\,A. Data points: amplitude of the anomalous magnetic field
variation $\Delta B_z$. Solid line: best fit of the modified
Bessel function $a K_1(2\pi(h+\delta)/\lambda)$. This has
$\lambda=217\pm10\,\mu$m and $\delta=251\pm12\,\mu$m. Dashed line:
The (usual) azimuthal field referred to the auxiliary ordinate on
the right.}
 \label{fig:decay}
\end{figure}

We see $\Delta B_z$ undergoing 2 full periods of oscillation along
the wire. The phase of the oscillation is fixed with respect to
the wire and did not change over several months of experiments.
The field has many more oscillations along the wire, with an
average wavelength $\lambda$ of $230\pm 10\,\mu$m, but in this
particular experiment the atoms are confined to these two periods
by the trap potential. It would seem that the current, or perhaps
the electron spin \cite{fortagh02,kraft02}, follows an oscillatory
or helical path along the wire with wavelength $\lambda$. This
could be due to some fundamental physics, or perhaps to a regular
arrangement of impurities or defects in the wire. In any such
case, the decay of the  field component $\Delta B_z$ with distance
$y$ should be well approximated by the modified Bessel function
$K_1(k y) \sim\exp(-ky)/\sqrt{ky}$, provided the distance $y$ from
the current to the atoms is much greater than the transverse
excursions of the current. The quantity $k$ is $2\pi/\lambda$. In
order to test this idea we have measured the amplitude of the
$\Delta B_z$ oscillations at each height $h$ above the wire and
these points are plotted in figure \ref{fig:decay}. The solid line
is a least-squares fit to the function $a
K_1(2\pi(h+\delta)/\lambda)$, in which $a, \delta$, and $\lambda$
were allowed to vary freely as fitting parameters. The best fit
has a $\chi^2$ of 4.8 for 5 degrees of freedom and gives the
results $\lambda=217\pm10\,\mu$m and $\delta=251\pm12\,\mu$m. The
close agreement between this value of $\lambda$ and the value of
$\lambda=230\pm10\,\mu$m obtained directly from the images
strongly suggests that our model is correct. The value of $\delta$
is equal to the wire radius indicating that the centre of the
oscillating current coincides with the centre of the wire.

 The dashed curve in figure \ref{fig:decay} shows
the usual magnetic field produced by 3.7\,A in a wire, which
varies inversely with the distance from the centre of the wire.
Expressed as a fraction of this field, the amplitude of the
$\Delta B_z$ oscillations ranges from $3\times 10^{-3}$ to
$3\times 10^{-4}$ over the heights studied. This ratio provides
information about the transverse excursion of the effective
current. For example, if we suppose that the current is helical
with an effective radius of $r$, the ratio of field components has
the value $k^2r y K_1 (k y)$, which implies that $r=50\,\mu$m. A
similarly large amplitude is found for other possible motions
(e.g. planar or random walk with characteristic length scale). The
Lorentz forces due to the applied magnetic fields do not generate
enough transverse current to account for this.  Nor do the
variations in the shape and centre of the wire. Microscopic
structure within the wire was revealed by polishing and etching a
longitudinal section of the wire following standard metallographic
methods. In the copper we found grains of $\sim10\,\mu$m and in
the aluminium we found small defects typically spaced by
$\sim5\,\mu$m. Neither of these provides a plausible explanation
for the transverse current. We therefore remain unable to propose
a mechanism for the observed variation of the current.

The exponential character of the decay means that $\Delta B_z$
decreases too rapidly with distance to be described by a power
law. The T\"{u}bingen group, who proposed a power law ansatz
 \cite{kraft02}, have kindly provided us with
their data \cite{fortaghthesis} taken above a 90\,$\mu$m diameter
copper wire and we find that it is also well described by the
$K_1$ Bessel function.

\section{Conclusions}

We have investigated the behaviour of a cloud of cold trapped
atoms brought close to the surface of a current-carrying wire. As
the atoms approach the wire, the cloud breaks into fragments due
to the presence of an anomalous magnetic field parallel to the
wire. Analysis of the density distribution of the atoms has
allowed us to make a quantitative map of this anomalous field.
Along the wire it oscillates with a period of 230$\pm 10\,\mu$m
and with increasing radius it decays according to a modified
Bessel function $K_1$. We point out that this behaviour is
characteristic of the field produced by a periodic transverse
current distribution in the wire. We have found a characteristic
decay length of $217\pm10\mu$m, in good agreement with the
observed period and we have shown that the anomalous current is
centred on the middle of the wire. Despite some effort, it remains
unclear why the current should move in this way.

 \ack
We are indebted to Alan Butler and John Knight for expert
technical assistance. We also thank J\'{o}zsef Fort\'{a}gh for
helpful discussions and for providing us with data from the
T\"{u}bingen experiment. This work was supported by the UK
Engineering and Physical Sciences Research Council, and by the
FASTNET and Cold Quantum Gases networks of the European Union.

\end{document}